\documentclass[usenatbib,usegraphicx]{mn2e}

\newcommand\nodata{\ldots}
\newcommand\ndet{\nodata & \nodata & \nodata}
\newcommand\ndetb{\nodata & \nodata & \nodata & \nodata }
\newcommand\nod{\nodata}
\newcommand\apjs{ApJS}
\newcommand\apj{ApJ}
\newcommand\aj{AJ}
\newcommand\aap{A\&A}
\newcommand\mnras{MNRAS}

\title[OH Maser Flares and Motions in W75N]{Flares and Proper Motions
  of Ground-State OH Masers in W75N}

\author[Fish, Gray, Goss, \& Richards]{
  Vincent L. Fish$^{1}$
  \thanks{E-mail: vfish@haystack.mit.edu (VLF); malcolm.gray@manchester.ac.uk (MG); mgoss@aoc.nrao.edu (WMG); a.m.s.richards@manchester.ac.uk (AMSR)},
  Malcolm Gray$^{2}$,
  W. M. Goss$^{3}$, and
  A. M. S. Richards$^{2}$\\
$^{1}$MIT Haystack Observatory, Route 40, Westford, MA 01886, USA\\
$^{2}$Jodrell Bank Centre for Astrophysics, The Alan Turing Building,
  School of Physics and Astronomy, The University of Manchester,\\
  Oxford Road, Manchester, M13 9PL, UK\\
$^{3}$National Radio Astronomy Observatory, 1003 Lopezville Rd.,
  P.O. Box O, Socorro, NM 87801, USA}

\begin{document}

\date{\today}

\pagerange{\pageref{firstpage}--\pageref{lastpage}} \pubyear{2011}

\maketitle

\label{firstpage}

\begin{abstract}
The star-forming region W75N hosts bright OH masers that are observed
to be variable.  We present observations taken in 2008 of the
ground-state OH maser transitions with the Very Long Baseline Array
(VLBA) and the Multi-Element Radio-Linked Interferometer Network
(MERLIN) and with the Nan\c{c}ay Radio Telescope in 2011.  Several of
the masers in W75N were observed to be flaring, with the brightest
1720-MHz maser in excess of 400~Jy.  The 1720-MHz masers appear to be
associated with the continuum source VLA~1, unlike the bright flaring
1665- and 1667-MHz masers, which are associated with VLA~2.  The
1720-MHz masers are located in an outflow traced by water masers and
are indicative of very dense molecular material near the \mbox{H\,{\sc
    ii}} region.  The magnetic field strengths are larger in the
1720-MHz maser region than in most regions hosting only main-line OH
masers.  The density falls off along the outflow, and the order of
appearance of different transitions of OH masers is consistent with
theoretical models.  The 1665- and 1667-MHz VLBA data are compared
against previous epochs over a time baseline of over 7 years.  The
median maser motion is 3.5~km\,s$^{-1}$ with a scatter that is
comparable to thermal turbulence.  The general pattern of maser proper
motions observed in the 1665- and 1667-MHz transitions is consistent
with previous observations.
\end{abstract}


\section{Introduction}

The Galactic star-forming region W75N, located at a distance of
1.3~kpc \citep{rygl2010}, is known to host bright OH masers with
significant variability.  Observations of the 1665-MHz masers have
detected flares above 1000~Jy, making W75N temporarily the brightest
OH maser source in the sky \citep*{alakoz2005,slysh2010}.  The
brightest masers, found near the continuum source VLA~2 \citep[as
  defined in][]{torrelles1997}, also exhibit the greatest variability.
Many observations have focused exclusively on the main-line 1665- and
1667-MHz transitions, which are distributed throughout W75N
\citep{fish2005}, rather than on the satellite-line 1720-MHz masers,
which are concentrated in one area and have previously been observed
to have a peak flux of only a few Jy \citep*{hutawarakorn2002}.

Owing to their intensity and compactness, masers are often used as
kinematic tracers of their environment, with motions usually being
deduced from multi-epoch very long baseline interferometric (VLBI)
observations.  \citet{norris1981}, \citet*{migenes1992}, and
\citet*{bloemhof1992} were among the first to measure proper motions
of OH masers in star-forming regions.  Due to the demanding nature of
VLBI observations, proper motion measurements of ground-state OH
masers have rarely been made over more than two epochs (e.g.,
\citealt*{wright2004}; \citealt{fishreid2007,fishreid2007b}).  In most
sources, these studies find large-scale ordered motions of the OH
masers, although individual motion vectors appear to have a large
random component.  One disadvantage of using a two-epoch set of
observations is the inability to determine what portion of an apparent
maser motion is real and what portion is due to random errors,
although the existence of large-scale ordered motions strongly
supports the interpretation of apparent motions as real, if noisy,
indicators of actual maser motions.  However, a recent multi-epoch
study of the ground-state OH masers in W75N by \citet{slysh2010}
suggests that the random errors dominate, preventing estimation of the
actual maser motions in most cases.

In this manuscript, we report on two epochs of observations of
ground-state OH transitions with the Very Long Baseline Array (VLBA)
and one epoch of observations of the 1720-MHz transition with the
Multi-Element Radio-Linked Interferometer Network (MERLIN).  These
observations occurred during a flare event in the 1720-MHz
transition.  In addition, we investigate the degree to which apparent
motions of OH masers are indicative of actual maser motions.

\section{Observations}

\subsection{VLBA}

W75N was observed with the VLBA on 2008 February 22 and 2008 April 26.
Total time on source was approximately 1.4~hr in February and 6.8~hr
in April.  The original purpose of the observations (experiment code
TC013) was to test and validate the output of the DiFX software
correlator against the Socorro hardware correlator, but the
observations were of suitable quality for scientific purposes as well.

The February observations focused on the 1665 and 1720~MHz transition,
while all four ground-state transitions of OH -- including 1612 and
1667~MHz -- were observed in April.  All data were taken in dual
circular polarization mode using 125-kHz bandwidths.  Data presented
in this manuscript were correlated with the Socorro correlator using
128 spectral channels.  These parameters correspond to an LSR velocity
span of 22~km\,s$^{-1}$ with a channel spacing of 0.18~km\,s$^{-1}$ at
1665~MHz.  Bandwidths were centred on LSR velocities of
6.0~km\,s$^{-1}$ at 1612 and 1665~MHz and 10.0~km\,s$^{-1}$ at 1667
and 1720~MHz.  The 1612-MHz data produced no detections and were
therefore not further analyzed.

Data were reduced using \textsc{aips}.  The source 3C345 was used as a
bandpass calibrator.  In each epoch, the data were calibrated using a
single spectral channel of a bright maser feature in 1665~MHz left
circular polarization (LCP) and another in 1720~MHz right circular
polarization (RCP). The resulting solutions were copied over to the
other polarization.  Data at 1667~MHz were phase-referenced to
1665~MHz.  Calibrated data were imaged using natural weighting and
1.5-mas pixel spacing.  The synthesized beamwidth was approximately $9
\times 6$~mas at a position angle near $-18\degr$ east of north, and
the blank sky, single-polarization thermal noise was $\sigma \approx
8$~mJy\,beam$^{-1}$ in the April epoch.

\subsection{MERLIN}

W75N was observed with MERLIN on 2008 May 13--14 in the 1720-MHz
transition only (rest frequency 1720.533~MHz).  Total on-source time
from 16 scans was 15 hours and 50 minutes. Dual circular polarization
observations covered a total bandwidth of 500~kHz in 256 spectral
channels with 0.34~km\,s$^{-1}$ spacing.  Channel 128 was set to an
LSR velocity of 8~km\,s$^{-1}$.

The phase calibrator, 2005+403, was observed nearly simultaneously
with the target but in a wide-band mode comprising a single 13-MHz
channel for greater sensitivity.  The flux calibrator 3C286 and
bandpass calibrator 3C84 were observed in both the narrow
spectral-line mode of the target and the wide mode of the phase
calibrator.

Data were reduced in \textsc{aips}.  After initial flagging of the
calibrator data, the task \textsc{calib} was run on both the narrow-
and wide-mode data for 3C84 in order to determine the phase offset
between these observing modes, which remained constant over the
duration of the observations.  The wide-mode phase-calibration table
was copied to the narrow-mode data and applied to correct for the
phase offset.  The narrow-band data for 3C84 were then calibrated in
phase and amplitude, using 17.47~Jy as the flux density in the
observing band.  The phase calibration consisted of a two-cycle
self-calibration of 2005+403 in wide mode.

The RCP data for W75N at 1720~MHz were self-calibrated using the
brightest channel.  The resulting solutions were then applied to LCP
to maintain the flux scale and the relative positional registration
between the polarizations.  The LCP data were subsequently
self-calibrated after zeroing the RCP solutions.  During processing,
data were weighted according to the sensitivity of each antenna.
Natural weighting was then used for imaging, and $1024 \times
1024$-pixel image cubes were generated.  Ten noisy channels were
dropped from each end of the frequency range.  The synthesized
beamwidth was $206 \times 142$~mas at a position angle of $-24\degr$,
and the blank sky, single-polarization thermal noise was
3~mJy\,beam$^{-1}$.

\subsection{Nan\c{c}ay}

Brief observations of the 1665- and 1720-MHz lines of OH were
conducted with the Nan\c{c}ay Radio Telescope on 2011 Jan 11.  Two
scans totalling 7.5~min were observed in dual linear and dual circular
polarization.  The observing bandwidth of 390.625~kHz was subdivided
into 1024 spectral channels, giving a resolution of 0.069~km\,s$^{-1}$
at 1665~MHz and 0.066~km\,s$^{-1}$ at 1720~MHz.  The observing band was
centred near $+26$~km\,s$^{-1}$ LSR.  Observations were frequency
switched with an offset of half the observing band, making the usable
velocity range approximately $-8$ to $+26$~km\,s$^{-1}$.  Data were
reduced using \textsc{class}.

\subsection{Identification of maser features}
\label{identification}

In each spectral channel of the AIPS and MERLIN data, maser components
detected above $7\,\sigma$ were fit with \textsc{aips} task
\textsc{orfit} (based on the standard \textsc{aips} task
\textsc{jmfit}) to determine their position, flux density, and size.
Maser components appearing in consecutive velocity channels and whose
centroid positions agree to within a fraction of the synthesized
beamwidth are identified as maser features.  The two circular
polarizations are analyzed independently.  For each maser feature,
Tables 1, 2, and 3 report the centroid position, flux density, and LSR
velocity of the channel of peak flux density.

The theoretical random error in determining the centroid position of a
maser component is approximately the synthesized beamwidth divided by
twice the signal-to-noise ratio \citep{reid88}, although for an
unfilled aperture this is an underestimate by a factor of a few
(\citealt*{richards1999}; see also \citealt{condon1998}).  Measurement
error in determining magnetic field strengths from Zeeman splitting is
proportional to the error in determining the velocity of peak emission
of the LCP and RCP components of each Zeeman pair, which in turn is
dominated by the velocity spacing of spectral channels
(0.176~km\,s$^{-1}$ at 1665 and 1667~MHz and 0.170~km\,s$^{-1}$ at
1720~MHz for the VLBA observations).  This imposes a quantization of
the measured magnetic field strengths in units of approximately 0.3,
0.5, and 1.5~mG at 1665, 1667, and 1720~MHz, respectively.

\subsection{Alignment of reference frames and flux conventions}
\label{alignment}

The VLBA observations of the 1667-MHz transition are phase-referenced
to the 1665-MHz transition.  The absolute position of the reference
feature at 1665~MHz is determined by identifying it with maser feature
A in \citet{slysh2010}, whose location was determined to be
$20^\mathrm{h}38^\mathrm{m}36\fs4082, +42^\circ37\arcmin34\farcs238$
(J2000) with an error of 2~mas in 2005 September.  At 1720~MHz, the
MERLIN observations obtained a position of the bright reference
feature.  This position is used to align the VLBA 1720-MHz data
relative to the 1665- and 1720-MHz positions.

Throughout this paper we report flux densities using the AIPS
convention ($I = (RR + LL)/2$).  The Nan\c{c}ay data are treated
similarly for consistency, although we note that this convention
differs by a factor of two from that which is sometimes used to report
Nan\c{c}ay data.

\section{Results}

\subsection{VLBA}

Counting LCP and RCP features separately, we identify 61 maser
features at 1665~MHz in the February data and 121 in the April data.
At 1720~MHz, we identify 20 features in the February data and 24 in
the April data, although spatial and spectral blending of the numerous
features in a small (approximately 125 $\times$ 350~mas) region makes
it difficult to uniquely and unambiguously identify some features.  In
the 1667-MHz transition, observed only in April, we identify 39 maser
features.

\begin{figure*}
\resizebox{\hsize}{!}{\includegraphics{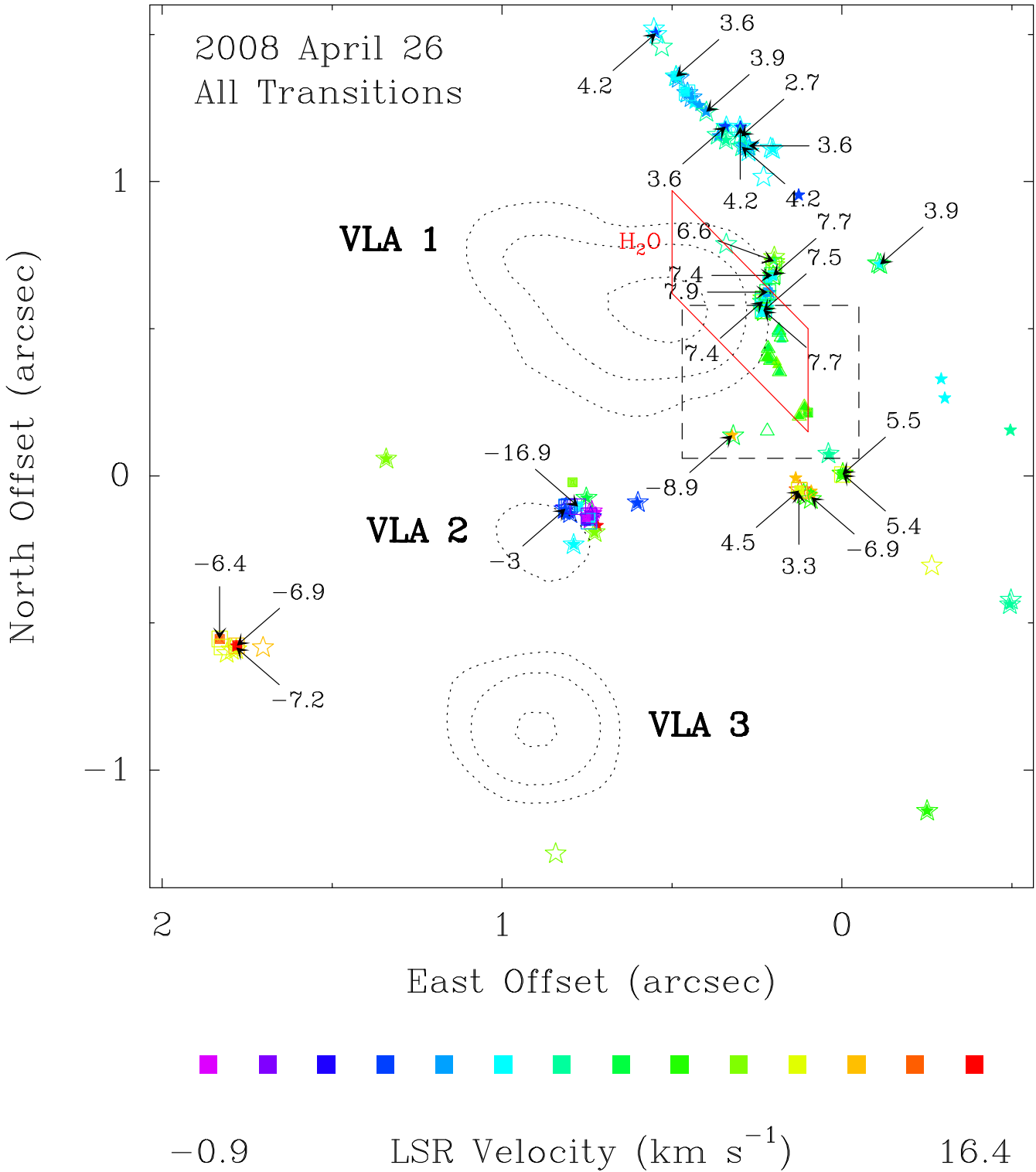}}
\caption{Image of detected OH masers in the 2008 April 26 data.
  Stars, squares, and triangles indicate 1665-, 1667-, and 1720-MHz
  masers, respectively.  LCP features are shown as filled symbols, and
  RCP features are shown as larger, open symbols, with colour used to
  indicate LSR velocity.  Not all known maser features in W75N fall
  within the velocity range of our observations
  \citep[see][]{fish2005}.  Numbers indicate magnetic field strengths
  in milligauss as determined from 1665- and 1667-MHz Zeeman pairs
  (see Fig.~\ref{fig-1720-map} for 1720-MHz Zeeman pairs).  Continuum
  emission at 8.4~GHz is shown in dotted contours
  (\citealt*{argon2000}; \citealt{fishreid2007}).  The position of the
  reference feature at the origin was measured in 2005 September to be
  $20^\mathrm{h}38^\mathrm{m}36\fs4082, +42^\circ37\arcmin34\farcs238$
  (J2000) \citep{slysh2010}.  All detected 1720-MHz masers appear in
  the region inside the dashed box, which is shown in greater detail
  in Fig.~\ref{fig-1720-map}.  The red parallelogram shows the region
  in which the highly variable water masers associated with VLA~1
  appear (\citealt{torrelles1997}; \citealt*{lekht2009}).
\label{fig-main-line-map}}
\end{figure*}

\begin{figure}
\resizebox{\hsize}{!}{\includegraphics{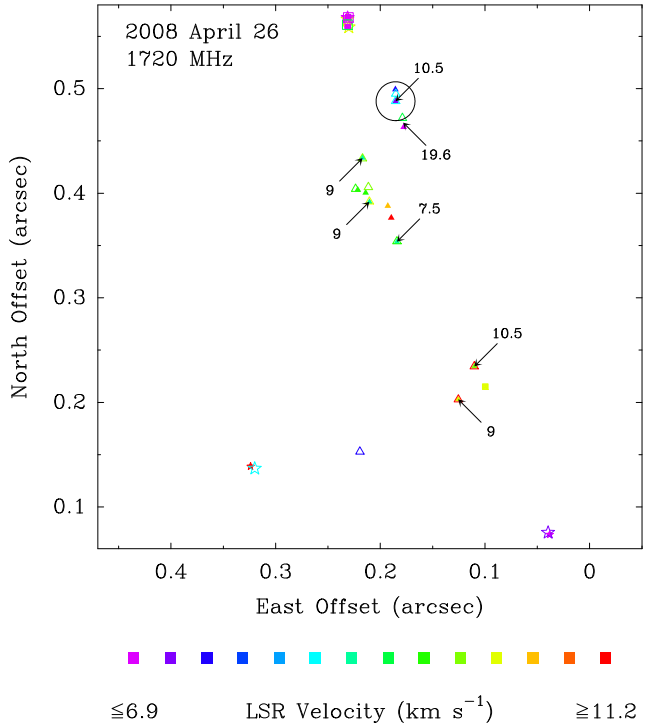}}
\caption{Enlargement of detected 1720-MHz OH masers in the 2008 Apil
  26 VLBA data.  The black circle denotes the location of the
  brightest maser.  The velocity scale is different from that of
  Fig.~\ref{fig-main-line-map}.  The origin is located at
  $20^\mathrm{h}38^\mathrm{m}36\fs4082, +42^\circ37\arcmin34\farcs238$
  (J2000), and the relative registration uncertainty between the
  1665/1667-MHz and the 1720-MHz features is $\sim 20$~mas.
\label{fig-1720-map}}
\end{figure}

\begin{figure}
\resizebox{\hsize}{!}{\includegraphics{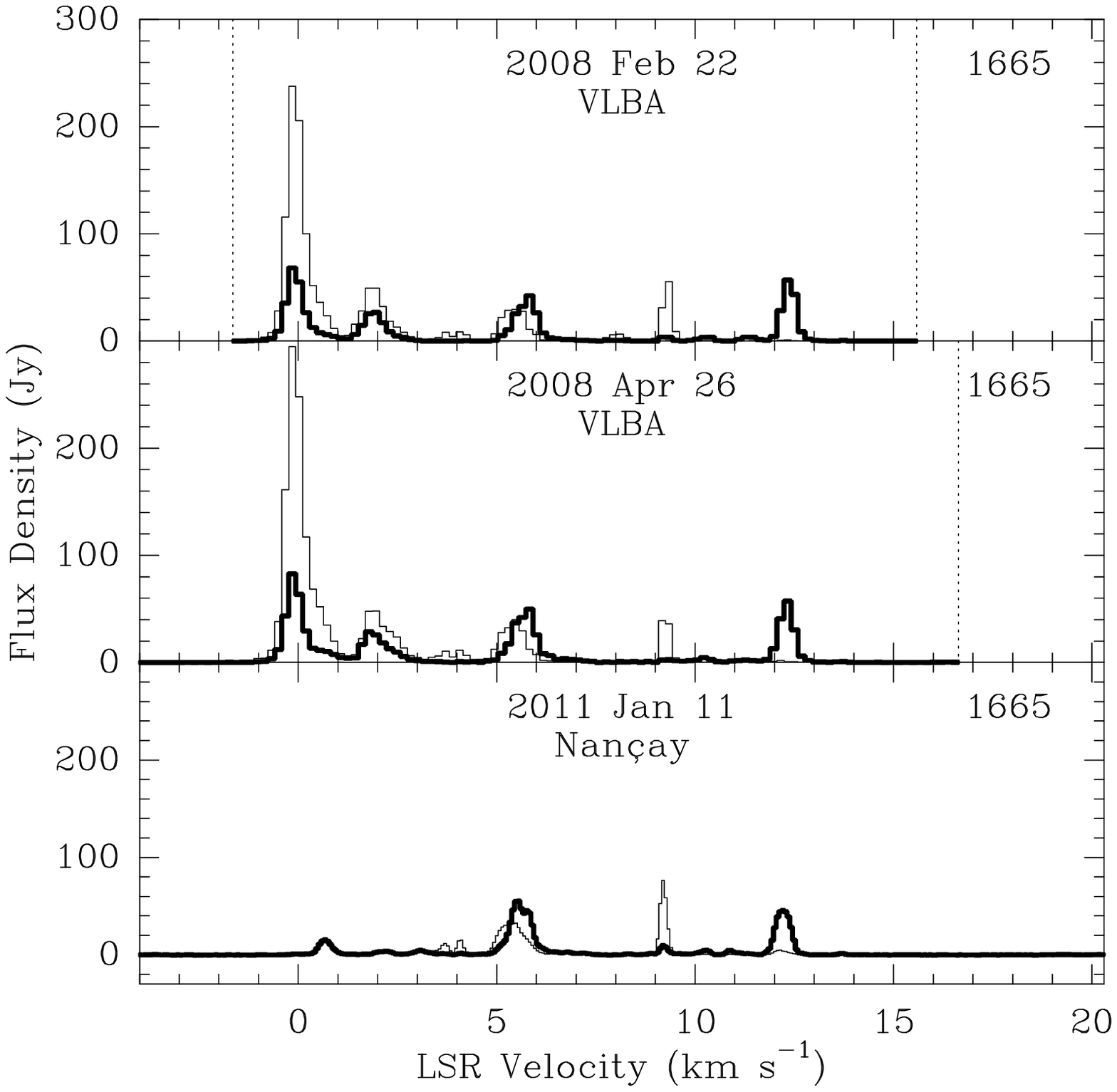}}
\resizebox{\hsize}{!}{\includegraphics{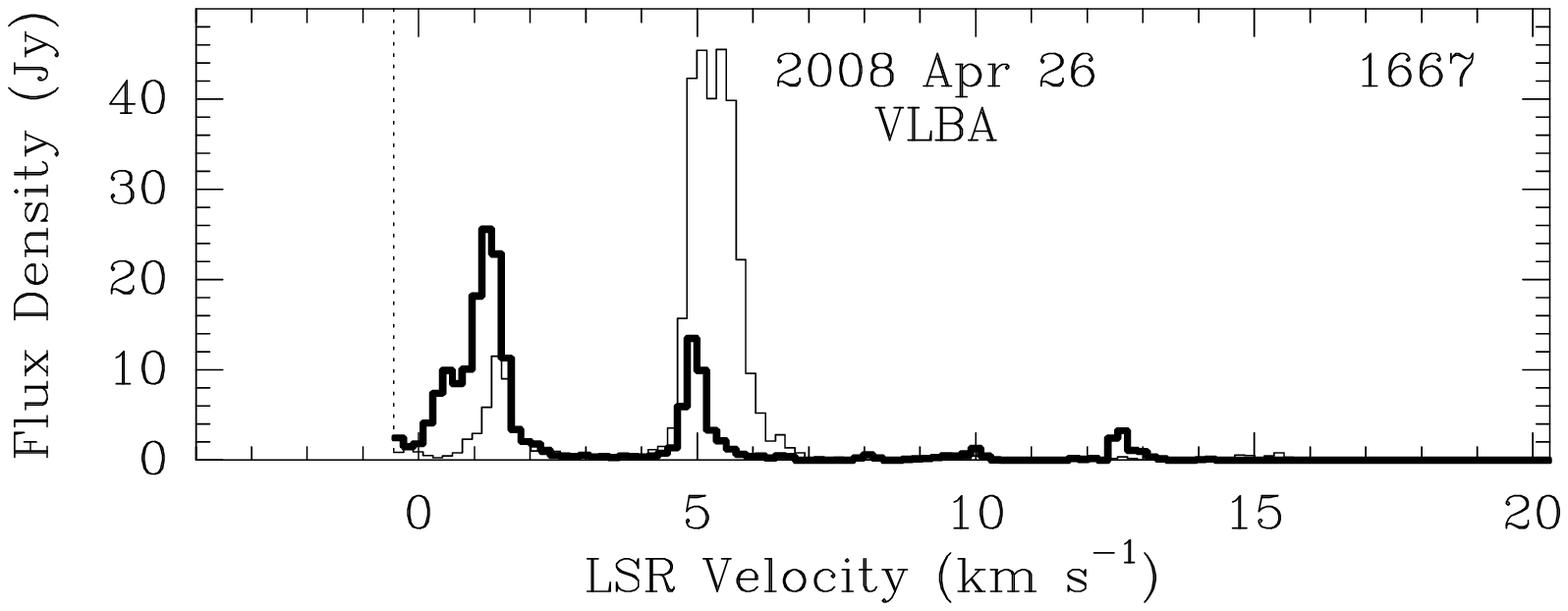}}
\resizebox{\hsize}{!}{\includegraphics{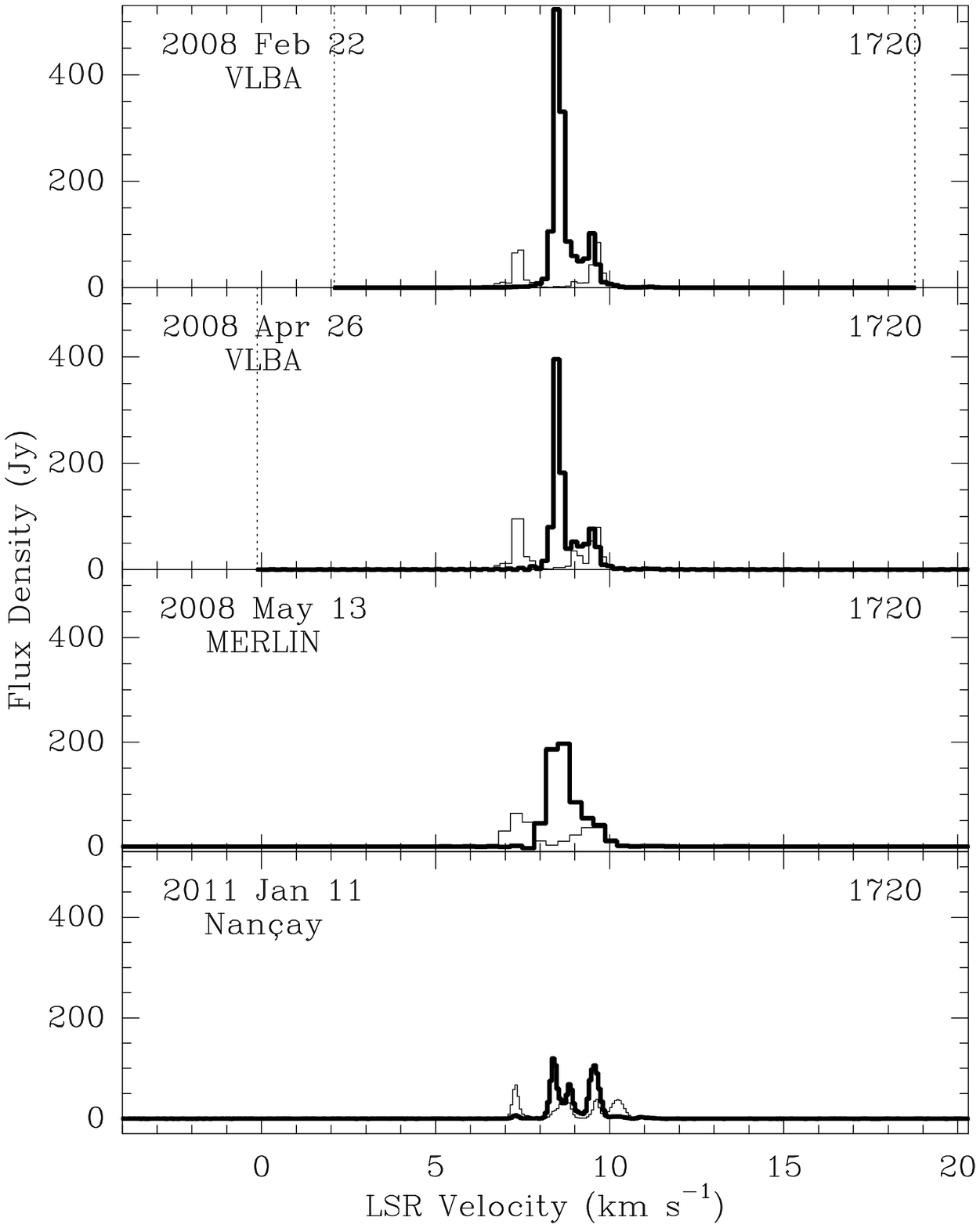}}
\caption{Spectra of the maser emission at 1665 (top), 1667 (middle),
  and 1720~MHz (bottom).  RCP and LCP emission are shown in bold and
  normal weight, respectively.  Dotted lines denote the last velocity
  channel imaged.  Spectra of the integrated maser emission from the
  May MERLIN data have coarser velocity resolution.
\label{fig-spectra}
}
\end{figure}

Parameters of the detected masers are given in Tables \ref{tab-1665},
\ref{tab-1667}, and \ref{tab-1720} for 1665-, 1667-, and 1720-MHz
masers, respectively.  Images of the April main-line and 1720-MHz data
are shown in Figs.~\ref{fig-main-line-map} and \ref{fig-1720-map}.
Spectra of the recovered emission (i.e., the total flux density in all
maser components as determined by fitting elliptical Gaussians) are
shown in Fig.~\ref{fig-spectra}.

The overall distribution of the 1665- and 1667-MHz maser features in
the 2008 epochs is very similar to those of the 2000/2001 and 2004
epochs \citep{fish2005,fishreid2007}.  Masers are located in an arc to
the north, an elongated cluster near the edge of VLA~1, several small
clusters near the origin in Fig.~\ref{fig-main-line-map}, a large
cluster on the northwest edge of VLA~2, and a smaller set of masers to
the east.  In addition, isolated maser spots appear throughout the
source, with many having counterpart identifications in previous
epochs.

Magnetic field strengths are also broadly consistent from epoch to
epoch, with the exception of the cluster near VLA~2.
\citet{slysh2010} note that the masers in this region exhibit
significant variability on time-scales of a few months.  Masers that
are detected in one epoch may no longer be detectable a few months
later, in which time new masers may have appeared.  The VLA-2 region
also contains numerous masers without an obvious Zeeman counterpart in
the opposite polarization.  Consequently, there are comparatively
fewer Zeeman magnetic field measurements in this region than in the
other clusters in W75N.  As in previous epochs
\citep{slysh2006,fishreid2007,slysh2010}, the largest main-line
magnetic field strength detected in these observations, approximately
17~mG, occurs in VLA~2.  (However, our narrow observing bandwidth does
not rule out the possibility of Zeeman pairs implying larger magnetic
field strengths.)

In addition, for the first time, we present a VLBI image of the
1720-MHz masers in W75N (Fig.~\ref{fig-1720-map}).  Unlike the
main-line masers, which are found throughout W75N, the 1720-MHz masers
are found only between the group of masers on the western edge of the
continuum emission from VLA~1 and the cluster of masers near the
origin of Fig.~\ref{fig-main-line-map} (to the southwest of VLA~1).
The magnetic fields deduced from Zeeman splitting at 1720~MHz are all
several milligauss higher than those derived from the 1665- and
1667-MHz transitions.  The line-of-sight direction of the magnetic
field at the 1720-MHz sites is oriented away from the observer
everywhere that it is detected, which is consistent with the direction
implied by the vast majority of the 1665- and 1667-MHz masers detected
north of the reference feature (at the origin in
Fig.~\ref{fig-main-line-map}).

Most of the 1720-MHz masers detected in one of the two 2008 epochs
were detected in the other and found to have similar intensity (to
within a few tens of per cent).  The differences between the two
epochs can largely be explained by the difficulty of identifying as
separate, maser components that are spatially blended (i.e., within a
synthesized beamwidth for most spots or more for the brightest spots,
where the noise is higher than thermal noise due to imperfect
calibration and deconvolution of the bright maser spot) or spectrally
blended (i.e., a secondary peak in the spectrum of a bright maser
located at roughly the same location but at a different velocity).

\subsection{MERLIN}
\label{MERLIN}

The MERLIN observations were made with the primary aim of determining
an accurate absolute position for the flaring maser object at
1720~MHz.  The position of the brightest (189~Jy\,beam$^{-1}$) RCP
maser feature, measured before self-calibration, was
$20^\mathrm{h}38^\mathrm{m}36\fs425, +42\degr37\arcmin34\farcs73$
(J2000).  The uncertainty in this position is dominated by effects
related to the phase calibration and has three main contributions:
$\sim 5$~mas from the uncertainties in the geographical positions of
the MERLIN telescopes, a contribution from the uncertainty in the
position of the phase calibrator itself (0.6~mas; L. Petrov, solution
rfc\_2010c, unpublished\footnote{available at
  http://astrogeo.org/vlbi/solutions/rfc\_2010c}), and a contribution
dependent on the angular separation between the phase calibrator and
target (8\degr), owing to the small difference in atmospheric
conditions between the two sources.  Assuming a positional uncertainty
of 2~mas\,degree$^{-1}$ from the latter effect, the astrometric
accuracy is 17~mas.

We can identify the brightest MERLIN feature, introduced above, with
the brightest RCP 1720-MHz feature from the April VLBA data (feature
12 in Table 3) with confidence.  At a typical maser velocity of
10~km\,s$^{-1}$, this feature would have moved a distance of less than
$0.1$~mas over the 18 days separating the April VLBA and MERLIN
observations.  The MERLIN position is used to align the 1720-MHz VLBA
reference frame with the 1665/1667-MHz frame, identifying the 1665-MHz
reference feature as feature~A in \citet{slysh2010}, whose position
was determined to 2-mas accuracy in 2005 September by
phase-referencing European VLBI Network (EVN) observations of the OH
masers to a nearby calibrator.

The flux densities of individual maser spots in the MERLIN data are
not directly comparable with those in the VLBA data due to the lower
spectral and spatial resolution.  However, the integrated flux
densities of all epochs can be compared.  The integrated Stokes I flux
density of the MERLIN data, 307~Jy\,km\,s$^{-1}$, is approximately the
same as measured in each of the VLBA epochs (310 and
271~Jy\,km\,s$^{-1}$ in February and April, respectively) to within
flux scale uncertainties.

\subsection{Variability}
\label{variability}

The most striking feature of the 1720-MHz data is the bright flare
producing a peak flux density exceeding 400~Jy (in an RCP
8.5~km\,s$^{-1}$ feature).  Bright features exceeding 60~Jy are also
detected at 7.3 and 9.7~km\,s$^{-1}$, with numerous other masers of a
few Jy or brighter detected over an LSR velocity range of $\sim
4$~km\,s$^{-1}$.  Flux densities were approximately persistent over a
period of three months in 2008 both on a spot-by-spot
(Table~\ref{tab-1720}) and an integrated flux density basis
(Section~\ref{MERLIN}).  The bright flare weakened between the 2008
MERLIN epoch and the Nan\c{c}ay observations in 2011.

In contrast, only three 1720-MHz maser features were detected with the
MERLIN array in 1993, with peak intensities of 3.7, 0.3, and
0.2~Jy\,beam$^{-1}$ \citep{hutawarakorn2002}.  We do not detect any of
these features, although the latter two may be below our sensitivity
limit.  The locations of these masers would be located near the
southern 1720-MHz masers in Fig.~\ref{fig-1720-map}.

Flares in W75N also occur in the 1665-MHz transition
\citep{alakoz2005}, with total emision from masers near 0 and
2~km\,s$^{-1}$ exceeding a flux density of 1000~Jy at their brightest
\citep{slysh2010}.  These masers are located near the continuum source
VLA~2.  In 2008, the 0~km\,s$^{-1}$ feature exceeded 200~Jy LCP and
the 2~km\,s$^{-1}$ feature exceeded 40~Jy LCP.  By 2011, the flux
densities of these features had dropped to the $\sim 1$~Jy level,
although a 15--20~Jy feature appeared near 0.7~km\,s$^{-1}$.  In
contrast, the flux densities of features with $v_\mathrm{LSR} \ga
3$~Jy are much more stable over the three years of observations.

At 1667~MHz, \citet{slysh2002} detected a 21~Jy feature at
9.8~km\,s$^{-1}$, which had weakened substantially by the epochs
reported on in \citet{slysh2010}.  The RCP feature near
15~km\,s$^{-1}$ brightened by nearly an order of magnitude between
2006 March and 2007 July but is not detected in our data (although a
much weaker LCP feature is detected).  We detect bright 1667-MHz
masers near 1.2~km\,s$^{-1}$ (25~Jy RCP, 11~Jy LCP) and 5~km\,s$^{-1}$
(44~Jy LCP, 13~Jy RCP), which correspond to weaker masers detected in
the 2007 July epoch of the \citet{slysh2010} Kalyazin RCP spectra but
not in the 2006 March or 2006 July epochs (although the 5~km\,s$^{-1}$
feature may be very weakly present in the latter), indicating that
these features were starting to brighten one year before our
observations.  Both the 1.2 and 5~km\,s$^{-1}$ features are located
near VLA~2.

\subsection{Comparison with previous VLBI data}
\label{previous}

Almost all of the masers reported by \citet{slysh2010} have
identifiable counterparts in our data.  At 1665~MHz, they detected 32
maser features in one of their 2006 epochs within the velocity range
of our 2008 April observations, excluding masers near VLA~2.  We find
corresponding masers in our data within a few milliarcseconds -- and
in almost all cases, within a few tenths of a km\,s$^{-1}$ -- of the
position and velocity given in Table~1 of \citet{slysh2010}.  We also
find masers within a few milliarcseconds of the locations of several
of the \citet{slysh2010} masers associated with VLA~2, although the
masers in this region are highly variable.  At 1667~MHz,
\citet{slysh2010} report on 12 masers in our velocity range, excluding
the VLA-2 region.  We find masers near 11 of these locations.

In the other direction, a substantial fraction of the masers we report
do not have counterparts listed in \citet{slysh2010}.  A direct,
one-to-one comparison of maser spots is made difficult by the
nonidentification of polarization information in the \citet{slysh2010}
data.  For instance, some of the masers we detect may correspond to
Zeeman counterparts of masers listed in the tables in
\citet{slysh2010}.  Other discrepancies may be due to methodological
differences in the data reduction.  Where masers are close enough
together for their emission to appear spatially blended at the
resolution of our array, we have attempted to fit two -- or rarely
more -- elliptical Gaussians to the emission.

\subsection{Magnetic fields}

The magnetic fields derived from the 1665- and 1667-MHz VLBA
observations are consistent in magnitude and line-of-sight direction
with values obtained in previous observations in all clusters of maser
spots, with the caveat that there is a large range of measured field
values near VLA~2 \citep{fish2005,fishreid2007}.  At 1720~MHz, we
detect several Zeeman pairs of approximately 9~mG as well as one
Zeeman pair indicating a field strength of about twice this value.

The magnetic field strengths we report at 1720~MHz are higher than the
magnetic field reported by \citet{hutawarakorn2002} for Zeeman pair
$Z_6$.  It is difficult to identify exact counterparts to the
\citet{hutawarakorn2002} detections in our VLBA data due to the much
lower angular resolution of the MERLIN data and significant flaring in
the intervening 15 years between the observations.  Nevertheless, we
do find a Zeeman pair at approximately the same velocity as
\citet{hutawarakorn2002} and with the same sense and approximate
magnitude of the velocity separation ($v_\mathrm{RCP} - v_\mathrm{LCP}
= 1.0$~km\,s$^{-1}$).  \citet{hutawarakorn2002} interpret this as due
to a $+4.6$~mG magnetic field, while we identify the field strength as
$+9$~mG.  The difference may arise due to uncertainty in the Zeeman
splitting coefficient for satellite-line masers.  \citet{davies1974}
provides the splitting coefficient appropriate if the LCP and RCP
features are due to a blend of the $\sigma^{\pm1}$, $\sigma^{\pm2}$,
and $\sigma^{\pm3}$ components in the intensity ratio expected in
local thermodynamic equilibrium.  However, very high spatial and
spectral resolution observations of the brightest 1612- and 1720-MHz
OH masers in W3(OH) uncovered no evidence for contributions of the
$\sigma^{\pm2}$ and $\sigma^{\pm3}$ components
\citep*{fishbrisken2006}.  The Zeeman splitting coefficient
appropriate for the $\sigma^{\pm1}$ components is half that
appropriate for the blend of all components, resulting in an estimate
of the magnetic field that is twice as large.

\subsection{Maser motions}

\begin{figure*}
\resizebox{0.49\hsize}{!}{\includegraphics{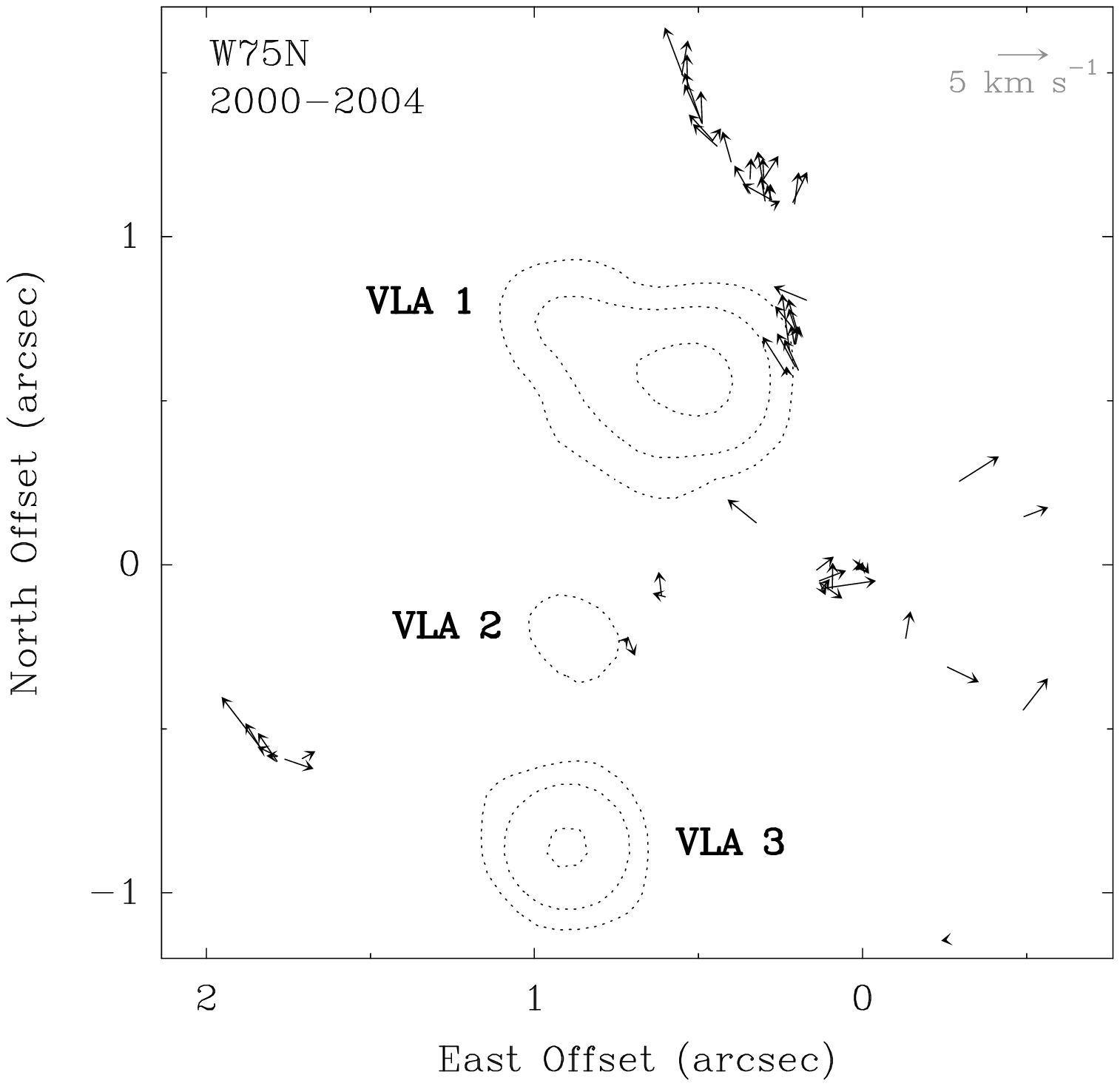}}
\resizebox{0.49\hsize}{!}{\includegraphics{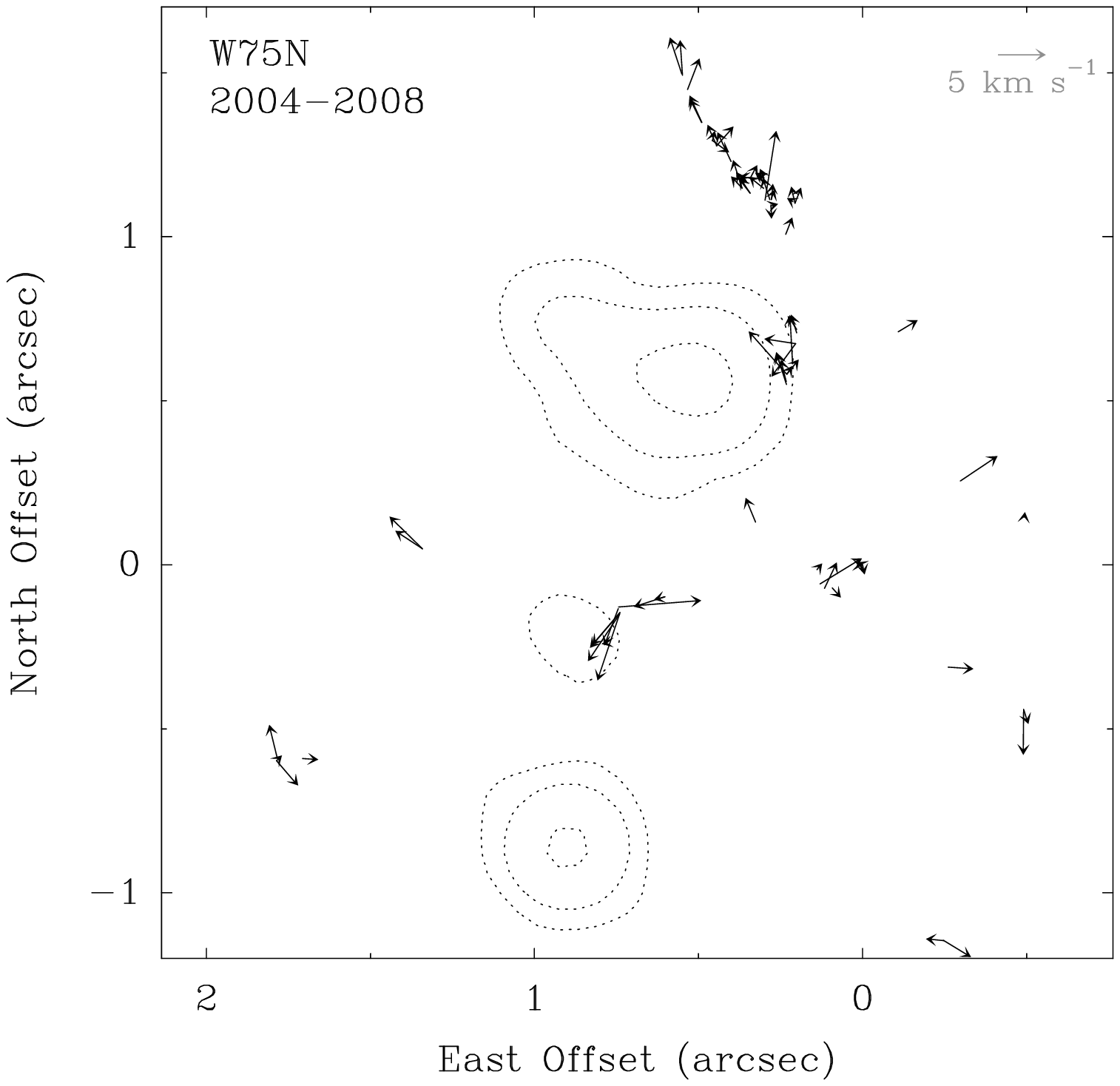}}
\caption{Apparent proper motions of masers in W75N between 2000 and
  2004 (\emph{left}) and between 2004 and 2008 (\emph{right}).  The
  left panel shows the proper motions from \citet{fishreid2007} in the
  reference frame of the 2008 data (i.e., keeping the 2008 reference
  feature stationary) after rescaling to a distance of 1.3~kpc
  \citep{rygl2010} but removing the misidentified 50~km\,s$^{-1}$
  motion \citep[see][]{slysh2010}.  A proper motion of 5~km\,s$^{-1}$
  corresponds to an apparent change in position of 0.8~mas\,yr$^{-1}$.
\label{fig-motions}
}
\end{figure*}

Fig.~\ref{fig-motions} shows the apparent relative maser motions
between the 2004 and 2008 epochs.  The reference frame has been chosen
such that the 1665-MHz reference feature at the origin is stationary.
The absolute motion of the reference feature cannot be determined from
our observations, since the VLBA data were not phase-referenced to a
nearby calibrator.

The main features in the apparent motions between 2004 and 2008 agree
with the main features in the motions between 2000/2001 and 2004
\citep{fishreid2007}.  Relative to the reference spot at the origin,
the northernmost masers are moving northward, possibly streaming along
the curve that outlines their general distribution.  The cluster of
masers at the western edge of the VLA-1 continuum region also show net
northward motion in both epochs.  The isolated masers found to the
west and south of the reference feature display net motion westward.
The general patterns of motions near VLA~2 and in the easternmost
cluster of masers are less clear.  We conclude that the general
pattern of motions is consistent with \citet{fishreid2007} and explore
the apparent differences statistically in Section~\ref{errors}.

\section{Discussion}

\subsection{Environment of the flaring 1720~MHz masers}

Strong flares in the 1665-MHz transition \citep{alakoz2005,slysh2010}
and weaker flares in the 1667-MHz transition
(Section~\ref{variability} of this work) are observed in the vicinity
of the continuum source VLA~2, a region where the OH masers are
rapidly variable and trace large magnetic fields
\citep{slysh2006,fishreid2007}.  In contrast, the bright 1720-MHz
flare on which we report is located $\sim 1000$~AU away from VLA~2.

It is difficult to say definitively whether the 1665/1667- and
1720-MHz flares are correlated, since the 1720-MHz transition has not
been monitored in W75N to nearly the same extent as the main-line
transitions.  If the flares are correlated, the flare excitation
process cannot be due to the passing of an MHD shock wave, which would
require $\sim 500$~yr to cross from one site to the other at
10~km\,s$^{-1}$.  Alternatively, the 1665/1667- and 1720-MHz flares
could be excited by a radiative mechanism, which can couple the region
in less than 6 days.  However, the correlation, if it exists, is not
very tight, since the 1665-MHz feature at 0~km\,s$^{-1}$ increases in
intensity, whilst the flaring 1720-MHz feature decreases slightly
between 2008 February and April (Fig.~\ref{fig-spectra}).  It is
therefore more plausible that the flaring events near VLA~1 and VLA~2
are unconnected.

Instead, the 1720-MHz OH masers are likely associated with the 22-GHz
water masers.  The 1720-MHz masers are located near the edge of the
continuum source VLA~1, aligned with water masers that lie along a
line oriented northeast-southwest through the continuum source
\citep{torrelles1997}.  Like the 1720-MHz OH masers, the 22-GHz H$_2$O
masers exhibit extreme variability.  Water masers appear and disappear
on time-scales of a year or less, with one feature reaching a flux
density of around 4000~Jy, becoming two orders of magnitude fainter in
two years, and then returning to over 1000~Jy a few years later
\citep{felli2007,lekht2009}.

\subsection{Models of the 1720~MHz masers}

\citet{torrelles1997} interpret the continuum emission from VLA~1 as
being due to a radio jet, noting that the presence of water masers
along this axis is consistent with this interpretation.  Based on the
lack of a clear velocity gradient in the LSR velocities of the water
masers along this axis, they speculate that the water masers are
shocked by the jet but not accelerated by it to any great degree,
suggesting that the water masers are embedded in dense clumps of
material.  The LSR velocities of the H$_2$O masers on the southeast
side of VLA~1 range from 10 to 16~km\,s$^{-1}$, which is close to the
velocity of the ambient material as traced by CS
\citep[10~km\,s$^{-1}$][]{hunter1994}.  The relatively small velocity
difference likely implies that the water masers reside near the
interface of the radio jet with the ambient material and are pumped by
the turbulent cascade of energy from the large-scale outflow to the
scale of individual water masers \citep{strelnitski2002}.

The properties of the OH masers fit this model.  The velocity range of
the 1720-MHz masers is 7 to 11~km\,s$^{-1}$, overlapping both the
velocity range of the water masers and the velocity of the ambient
material.  The magnetic fields measured by 1720-MHz Zeeman splitting,
up to 20~mG, are about three times larger than usual for OH masers in
star-forming regions, suggesting that the material that they trace is
a factor of $\sim 10$ denser under the usual assumption that density
is correlated with magnetic field strength \citep[$B \propto
  n^\alpha$, $\alpha \approx 0.5$;][]{crutcher1999}.  The existence of
flaring OH masers in the 4765-MHz transition in this region
\citep{niezurawska2004} also supports a high density.
\citet*{gray1992} find that strong 1665-MHz maser emission can arise
at $n_\mathrm{OH} \approx 20$~cm$^{-3}$, while at $n_\mathrm{OH} \ge
200$~cm$^{-3}$ strong 1720- and 4765-MHz emission is seen without
accompanying 1665-MHz emission (see their Figure 12).  The existence
of a substantial region of parameter space where 1720- and 4765-MHz
masers co-exist at OH specific column densities that are too high for
1665- and 1667-MHz masers is also supported by \citet*{cragg2002}.  On
the assumption that the number density of OH is a good proxy for the
total number density, these models are also consistent with the ratio
of densities in the two masing regions as derived from the magnetic
field strengths.

The distribution of 1720-MHz OH masers along the line of water masers
argues in favour of the 1720-MHz OH masers being associated with an
outflow rather than a disc.  In this regard, the outflow model of
W3(OH) \citep{gray1992} may be more directly applicable to the
1720-MHz masers in W75N than the disc model of \citet*{gray2003}.  The
W3(OH) model produces 1720- and 4765-MHz OH maser emission near the
base of an outflow and 1665-MHz emission farther out, which is
observationally consistent with the location of OH masers in W75N
along the outflow near VLA~1 as well.  However, an important
difference between W3(OH) and W75N is that in the former the 1720-MHz
masers are intermixed with masers at 1665~MHz, while in the latter the
1720-MHz region is devoid of 1665-MHz masers.  More quantitatively,
there exist 1665--1720~MHz pairs separated by only a few AU in W3(OH)
\citep*{fishbrisken2006}, while the separation of the closest such
pair in W75N exceeds 100~AU.  For comparison, there exists a weak
1667-MHz maser located $\sim 30$~AU from the nearest 1720-MHz maser in
W75N.

The nonoverlap of 1665- and 1720-MHz masers in W75N may be explained
by a larger density than in W3(OH).  In a slowly accelerating flow
(velocity gradient of 1000~km\,s$^{-1}$\,pc$^{-1}$, or $\sim
0.3$~km\,s$^{-1}$ per $10^{13}$~m), competitive gain produces strong
1720-MHz emission while quenching 1665-MHz emission \citep{gray1992}.
W75N is believed to be younger than W3(OH), and the density of
molecular material near the \mbox{H\,{\sc ii}} region is likely to be
higher, making competitive gain effects more important in W75N than in
W3(OH) \citep[see discussion in][]{gray2003}.  Along the direction of
the flow, the first ground-state maser to appear beyond the end of the
1720-MHz maser zone is a weak 1667-MHz maser, followed much farther
out by a weak 1665-MHz maser and then strong 1665-MHz maser emission
(the reference feature).  The magnetic field strengths falls off from
about 9~mG at the end of the 1720-MHz region to 5.5~mG at the location
of the bright 1665-MHz emission farther along the flow, indicating
that the density decreases by about a factor of 3 away from the
\mbox{H\,{\sc ii}} region.  These observed properties are consistent
with the various Gray et al.\ models, which predict that as the gain
length increases, 1720-MHz masers will be produced while first 1665-
and then 1667-MHz masers are quenched.

\subsection{Measurement of proper motions}
\label{errors}

The recent six-epoch data set of \citet{slysh2010} highlighted the
importance in estimating the magnitude of random errors in determining
the positions -- and hence apparent proper motions -- of OH masers in
star forming regions.  Previous proper motion measurements of 1.6-GHz
masers based on two epochs of data (e.g., \citealt{fishreid2007} for
W75N and \citealt{bloemhof1992} and \citealt{fishreid2007b} for other
sources) do not have the ability to discriminate between true maser
motions and non-real apparent motions caused by both random (e.g.,
thermal noise) and systematic (e.g., deconvolution) errors
(Section~\ref{identification}).  An additional complication is that
maser proper motions are typically assumed to be linear, yet real
masers may undergo accelerations that produce nonlinear trajectories.
Turbulence may further complicate interpretation of apparent proper
motions.

\citet{slysh2010} demonstrate that the total of these effects can be
much larger than what would be expected from the random uncertainty
due to thermal noise alone.  The error in determining the maser
position can be dominated by other effects, such as calibration and
deconvolution errors or changes in the maser substructure on spatial
scales smaller than the VLBA resolution of $\sim 7$~mas at 1.6~GHz
\citep{fishsjouwerman2007}.  A discussion of the difficulty in
identifying the ``peak'' of a maser in a different context can be
found in \citet*{moellenbrock2009}.  On the other hand, the
reproducibility of the general features of the 1665-MHz OH proper
motions both in W3(OH) \citep{wright2004} and in W75N (this work,
Fig.~\ref{fig-motions}) suggests that apparent maser proper motions,
when viewed as an ensemble, are reliable estimators of motions of
maser clusters provided that a sufficiently long time baseline has
elapsed.

One estimate of these errors can be obtained by comparing the
positions of masers in the 2008 February and April data.  At 1665 MHz,
54 features are detected in both the February and April data.  Of
these, all but 6 have centroids that agree to better than 1~mas
(equivalently, a false apparent proper motion of 36~km\,s$^{-1}$)
between the two epochs, with a median position difference of 0.4~mas
(15~km\,s$^{-1}$) after referencing to the same bright maser spot.
The position differences between the 2008 February and April epochs
provide an estimate of the total error in determining maser proper
motions.  These are not indicative of true proper motions, since the
implied velocities far exceed those measured between 2000 and 2004
\citep{fishreid2007} and between 2004 and 2008 April (this work;
median velocity 3.5~km\,s$^{-1}$).

\begin{figure}
\resizebox{0.49\hsize}{!}{\includegraphics{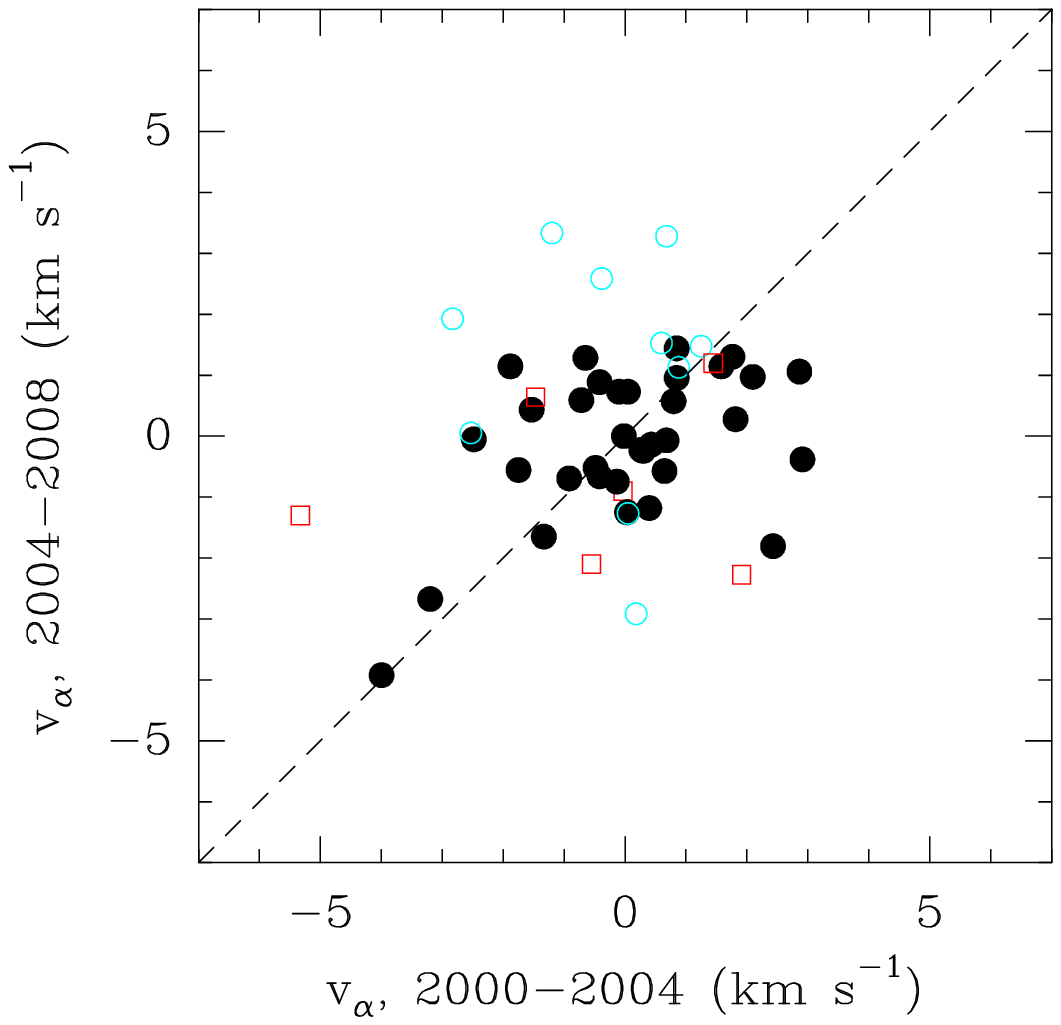}}
\resizebox{0.49\hsize}{!}{\includegraphics{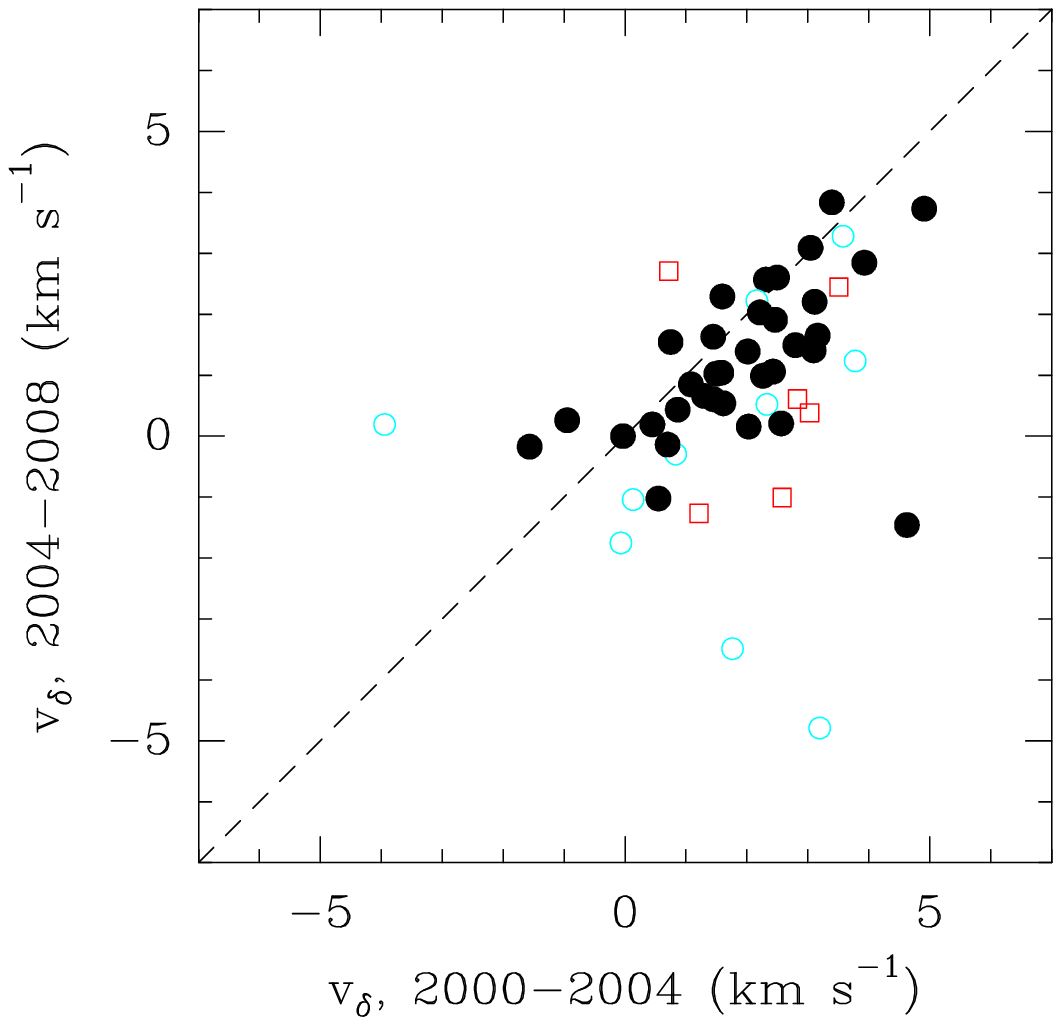}}
\caption{Comparison of apparent maser proper motions between epochs.
  The RA (\emph{Left}) and Dec (\emph{Right}) component of motions
  between the 2004 and 2008 April epochs versus the same component
  between the 2000 and 2004 epochs.  Spots that are spatial blends of
  emission from two masers in the 2008 April epoch are shown as open
  red squares.  Open cyan circles denote masers whose minimum flux
  density in the three epochs is less than 0.2~Jy.  Bright unblended
  masers are shown as filled black circles.  The correlation between
  epochs of proper motions obtained from unblended masers indicates
  that the apparent proper motions are real.
\label{fig-motion-errors}
}
\end{figure}

A second way to determine whether apparent proper motions are
indicative of real maser motions is to compare the apparent proper
motions across more than one epoch.  To the extent that the apparent
maser motions are real and linear, the apparent motions between epochs
1 and 2 should be equal to the apparent motions between epochs 2 and
3, with the scatter providing a measurement of the total error
associated with measuring proper motions and the effects of
turbulence.  We select the three epochs (1) 2000 November 22 and 2001
January 6 \citep{fish2005}, (2) 2004 September 16/19
\citep{fishreid2007}, and (3) 2008 April 26 (this work).  These epochs
are roughly equally spaced in time, were all observed with several
hours of on-source observing time with the same array (VLBA), and were
all reduced and analyzed in a similar manner.  (Other epochs of VLBI
data exist, such as the VLBA snapshot of \citet{slysh2002} and the
three EVN epochs of \citet{slysh2010}, but the three selected epochs
have higher sensitivity (see Section~\ref{previous}).)
Fig.~\ref{fig-motion-errors} shows the correlation between proper
motions from epoch 1 to 2 and from epoch 2 to 3.  Excluding masers
that are weaker than 0.2~Jy in an epoch (for which signal-to-noise
limitations in determining the centroid position in each epoch
dominate) or that are spatially blended in the third epoch, the rms
velocity spread in each of right ascension and declination is
1.1~km\,s$^{-1}$, with median values lower by a factor of two.  In
comparison, detected proper motions are typically several times this
value (Figure~\ref{fig-motions}), suggesting that the apparent proper
motions are real.  We also note that the velocity spread is very close
to the intrinsic thermal linewidth of water masers near VLA~1
\citep{surcis2011} and comparable to the estimate of the turbulent
velocity in OH maser cluster in W3(OH) \citep{reid1980}, possibly
indicating that turbulence is partially responsible for the observed
scatter.

A third, though less robust, estimator of the noise in apparent maser
proper motions is the magnitude of the vector difference of velocities
between the epochs, $|v_{23} - v_{12}|$, where the subscripts indicate
epoch numbers.  Neglecting weak and spatially blended masers, the
median value of $|v_{23} - v_{12}|$ is 1.5~km\,s$^{-1}$.  However,
this estimator is not robust against random errors in determining the
position of the reference feature in each epoch.  For instance, a
0.5-mas error in the position of the reference feature in the third
epoch could lower the median vector velocity difference to below
1.1~km\,s$^{-1}$.  (In contrast, a positional error in the reference
feature shifts points in the panels in Fig.~\ref{fig-motion-errors}
without affecting the scatter among the points.)

These statistics are consistent with the result of \citet{slysh2010}
that the errors in determining OH maser proper motions are
significantly larger than would be estimated from the errors in
determining maser centroids due solely to signal-to-noise
considerations (at least above a detection threshold of $\sim
20\,\sigma$).  Nevertheless, it is also apparent that the apparent
proper motions of OH masers are indeed tracing their local velocity
field.  Proper motion maps of OH masers can be a reliable estimator in
determining motions of clusters containing multiple bright maser spots
that are spatially unblended.  However, caution should be exercised in
interpreting small ($\la 0.1$~beamwidth) motions of individual
masers, where systematic errors effectively introduce a large error in
determining the position angle of the motion.

\section{Conclusions}

We detect a strong flare in the 1720-MHz OH maser emission in W75N,
with the brightest RCP feature exceeding a flux density of 400~Jy.
Although weak 1720-MHz masers have been detected in W75N before, the
observations we report are two orders of magnitude brighter than
previously observed.  The 1720-MHz masers are likely produced by very
dense molecular material that is excited and slowly accelerated by the
outflow traced by highly-variable water masers near the continuum
source VLA~1.  The appearance of main-line masers implying smaller
magnetic fields farther down the line of the outflow support
theoretical OH maser models, which predict 1720- and 4765-MHz masers
at higher densities than 1665- and 1667-MHz masers.

When compared with VLBI images of the 1665- and 1667-MHz OH masers in
previous epochs, the VLBA observations reported herein are consistent
with proper motions previously reported, contrary to the findings of
\citet{slysh2010}.  We conclude that apparent maser motions are
reliable estimates of real proper motions but note that the
measurement error in determining proper motions is larger than would
be expected from signal-to-noise considerations alone.

\section{Acknowledgments}

The National Radio Astronomy Observatory is a facility of the National
Science Foundation operated under cooperative agreement by Associated
Universities, Inc.  MERLIN is a National Facility operated by the
University of Manchester at Jodrell Bank Observatory on behalf of
STFC.  We thank Pierre Colom for arranging the observations at
Nan\c{c}ay.

\begin{table*}
\begin{minipage}{175mm}
\caption{Detected Masers at 1665~MHz}
\label{tab-1665}
\begin{tabular}{rrrrrrrrrr}
\hline
\multicolumn{3}{c}{April 2008} & & & \multicolumn{3}{c}{February 2008} & 2004 & 2000 \\
{RA Off.} &
{Dec Off.} &
{$S_\nu$} &
{$v_\mathrm{LSR}$} &
{} &
{RA Off.} &
{Dec Off.} &
{$S_\nu$} &
{Feature} &
{Feature} \\
{(mas)} &
{(mas)} &
{(Jy)} &
{(km\,s$^{-1}$)} &
{Pol} &
{(mas)} &
{(mas)} &
{(Jy)} &
{Number} &
{Number} \\
\hline
$-$494.89 &$-$428.70 &   0.08 & 7.41 & R & \ndet                         &\nod &\nod \\
$-$493.42 &   148.18 &   0.23 & 6.88 & L & \ndet                         &   1 &   1 \\
$-$492.16 &$-$444.00 &   0.06 & 7.41 & R & \ndet                         &   2 &\nod \\
$-$491.52 &$-$443.40 &   0.09 & 7.41 & L & \ndet                         &   2 &   2 \\
$-$300.55 &   257.63 &   1.22 & 6.00 & L & $-$300.56 &   257.58 &   1.11 &   3 &   3 \\
$-$289.50 &   322.25 &   0.31 & 6.00 & L & \ndet                         &\nod &\nod \\
$-$262.22 &$-$312.48 &   0.84 &12.68 & R & $-$262.51 &$-$313.01 &   0.76 &   4 &   4 \\
$-$248.49 &$-$1145.77&   0.18 & 9.69 & R & \ndet                         &   5 &   5 \\
$-$248.11 &$-$1145.40&   0.14 & 9.69 & L & \ndet                         &   5 &   5 \\
$-$110.40 &   711.29 &   0.14 & 8.29 & R & \ndet                         &   8 &\nod \\
$-$109.82 &   711.62 &   0.20 & 6.00 & L & \ndet                         &\nod &\nod \\
$-$103.41 &   714.99 &   0.22 & 8.29 & R & \ndet                         &\nod &\nod \\
     0.00 &     0.00 &  45.10 &12.33 & R &      0.00 &     0.00 &  44.95 &  10 &   7 \\
     0.02 &  $-$0.22 &   1.95 & 9.16 & R &   $-$0.01 &  $-$0.25 &   3.15 &   9 &   8 \\
     0.10 &  $-$0.02 &  36.38 & 9.16 & L &   $-$0.01 &  $-$0.33 &  50.26 &   9 &   8 \\
     \ndetb                          & L &      0.10 &  $-$0.11 &   0.69 &\nod &\nod \\
    40.47 &    66.35 &   0.44 & 7.05 & L &     39.12 &    64.63 &   0.60 &\nod &\nod \\
    42.05 &    68.20 &   0.14 & 7.23 & R & \ndet                         &\nod &\nod \\
    94.00 & $-$86.65 &   0.15 & 9.69 & R & \ndet                         & (13)&\nod \\
    91.17 & $-$71.43 &   0.26 &13.73 & L & \ndet                         &  12 &  10 \\
    92.29 & $-$59.07 &   0.06 &13.73 & L & \ndet                         &\nod &\nod \\
   106.58 & $-$73.65 &   0.12 & 9.69 & R & \ndet                         &\nod &\nod \\
   114.94 & $-$69.53 &   0.29 & 9.34 & R & \ndet                         &  14 &  11 \\
   126.45 & $-$56.80 &   0.06 &14.61 & R & \ndet                         &  17 &\nod \\
   129.48 &   947.26 &   0.36 & 3.36 & L &    129.80 &   947.71 &   0.44 &\nod &\nod \\
   131.03 & $-$67.10 &   0.25 &12.68 & L & \ndet                         &\nod &  15 \\
   138.51 & $-$14.11 &   0.24 &12.86 & L & \ndet                         &  19 &  16 \\
   201.13 &   718.81 &   0.40 &11.10 & R & \ndet                         &  25 &  21 \\
   201.27 &   708.55 &   0.80 &10.92 & R & \ndet                         &  26 &  22 \\
   201.51 &   736.57 &   0.11 &11.10 & R & \ndet                         &\nod &\nod \\
   204.16 &   726.41 &   0.08 & 7.23 & L & \ndet                         &\nod &\nod \\
   204.82 &   671.94 &   0.08 & 8.99 & R & \ndet                         &  29 &  23 \\
   205.25 &   674.25 &   0.08 & 4.42 & L & \ndet                         &  28 &  24 \\
   205.95 &  1103.09 &   5.84 & 5.47 & R &    206.26 &  1103.30 &   2.52 &  31 &  26 \\
   206.06 &  1103.49 &   0.74 & 5.47 & L &    206.80 &  1103.75 &   0.62 &  31 &\nod \\
   212.00 &  1107.10 &   1.04 & 5.82 & R & \ndet                         &  32 &  27 \\
   213.15 &   616.32 &   0.22 & 4.59 & L & \ndet                         &  33 &\nod \\
   229.76 &   582.15 &   5.67 & 5.30 & L &    229.69 &   582.21 &   6.81 &  35 &  29 \\
   230.30 &   581.80 &   1.17 & 5.47 & R &    230.24 &   582.47 &   0.57 &  35 &  29 \\
   232.55 &   551.53 &   0.65 &10.04 & R & \ndet                         &  37 &\nod \\
   233.08 &  1008.85 &   0.36 & 5.30 & R & \ndet                         &  39 &\nod \\
   233.57 &   560.54 &  16.42 & 5.47 & L &    233.48 &   559.86 &  12.79 &  38 &\nod \\
   233.59 &   560.47 &   2.38 & 5.47 & R &    233.43 &   559.65 &   2.23 &  38 &\nod \\
   233.69 &   596.00 &   1.98 &10.39 & R &    234.04 &   595.80 &   3.17 &\nod &\nod \\
   233.78 &   559.07 &   0.78 &10.04 & R & \ndet                         &\nod &\nod \\
   234.13 &   584.26 &   0.38 &10.22 & R & \ndet                         &  36 &  28 \\
   234.38 &   588.46 &   1.45 & 5.82 & L & \ndet                         &\nod &\nod \\
   276.55 &  1113.87 &   0.55 & 4.42 & L & \ndet                         &\nod &\nod \\
   277.38 &  1114.97 &   0.70 & 6.53 & R & \ndet                         &  40 &  30 \\
   277.49 &  1094.61 &  11.83 & 5.82 & R &    277.84 &  1094.92 &   8.20 &  41 &  32 \\
   278.63 &  1113.82 &   0.36 & 4.24 & R &    278.81 &  1113.88 &   0.24 &  42 &  31 \\
   279.07 &  1096.29 &   2.93 & 5.65 & L &    279.43 &  1096.96 &   2.19 &  41 &  32 \\
   282.93 &  1114.59 &   8.57 & 4.07 & L &    282.90 &  1114.58 &   5.20 &  42 &  33 \\
   289.34 &  1115.25 &   0.28 & 5.82 & R & \ndet                         &\nod &\nod \\
   296.55 &  1111.64 &   1.51 & 6.88 & R &    296.41 &  1111.93 &   0.96 &  43 &  34 \\
   296.68 &  1144.82 &   0.18 & 5.65 & L & \ndet                         &\nod &\nod \\
   297.11 &  1111.36 &   0.11 & 4.42 & L & \ndet                         &\nod &\nod \\
   300.52 &  1178.84 &   0.44 & 3.19 & L &    300.08 &  1178.47 &   0.48 &  44 &  36 \\
   301.82 &  1148.25 &   0.05 & 7.23 & R & \ndet                         &  45 &  35 \\
   302.06 &  1178.36 &   0.47 & 5.65 & R & \ndet                         &  46 &  37 \\
   302.10 &  1178.44 &   0.73 & 5.65 & L &    302.39 &  1178.58 &   0.87 &  46 &  37 \\
   322.37 &   129.46 &   0.16 & 8.64 & R & \ndet                         &\nod &\nod \\
   326.40 &   131.51 &   0.64 &13.91 & L &    326.61 &   131.73 &   0.76 &  47 &  38 \\
\end{tabular}
\end{minipage}
\end{table*}
\begin{table*}
\begin{minipage}{175mm}
\begin{tabular}{rrrrrrrrrr}
\hline
\multicolumn{3}{c}{April 2008} & & & \multicolumn{3}{c}{February 2008} & 2004 & 2000 \\
{RA Off.} &
{Dec Off.} &
{$S_\nu$} &
{$v_\mathrm{LSR}$} &
{} &
{RA Off.} &
{Dec Off.} &
{$S_\nu$} &
{Feature} &
{Feature} \\
{(mas)} &
{(mas)} &
{(Jy)} &
{(km\,s$^{-1}$)} &
{Pol} &
{(mas)} &
{(mas)} &
{(Jy)} &
{Number} &
{Number} \\
\hline
   342.21 &  1178.75 &  19.47 & 5.47 & R &    342.15 &  1178.77 &  14.21 &  48 &  40 \\
   342.67 &  1178.55 &   2.38 & 5.65 & L &    342.29 &  1178.34 &   2.45 &  48 &  40 \\
   342.82 &   780.47 &   0.08 & 6.88 & R & \ndet                         &\nod &\nod \\
   343.76 &  1134.23 &   0.31 & 6.53 & L &    344.21 &  1134.24 &   0.30 &  49 &  39 \\
   343.78 &  1134.11 &   1.40 & 6.53 & R &    343.91 &  1134.28 &   1.27 &  49 &  39 \\
   346.84 &  1180.12 &   0.68 & 3.36 & L &    346.49 &  1180.49 &   0.54 &  50 &\nod \\
   366.68 &  1148.67 &   0.33 & 4.59 & L &    366.55 &  1148.14 &   0.37 &  51 &\nod \\
   368.15 &  1151.27 &   0.11 & 7.05 & R & \ndet                         &  52 &\nod \\
   401.26 &  1229.21 &   0.06 & 7.05 & R & \ndet                         &\nod &\nod \\
   401.75 &  1231.33 &   0.42 & 4.77 & L &    401.65 &  1230.54 &   0.45 &  53 &  41 \\
   421.41 &  1251.55 &   0.17 & 5.12 & L & \ndet                         &\nod &\nod \\
   439.11 &  1258.69 &   0.29 & 5.47 & L & \ndet                         &\nod &\nod \\
   443.98 &  1278.14 &   1.61 & 5.12 & L &    444.22 &  1278.27 &   1.30 &  54 &  42 \\
   444.96 &  1278.65 &   0.31 & 5.12 & R & \ndet                         &  54 &  42 \\
   456.51 &  1292.55 &   5.93 & 5.12 & L &    457.75 &  1293.74 &   6.18 &  55 &  43 \\
   457.01 &  1293.31 &   3.42 & 5.12 & R &    457.25 &  1293.31 &   2.88 &  55 &  43 \\
   460.53 &  1296.87 &   3.61 & 5.30 & L & \ndet                         &\nod &\nod \\
   482.67 &  1344.21 &   0.14 & 4.95 & R & \ndet                         &  56 &\nod \\
   482.83 &  1344.49 &   2.07 & 4.95 & L &    482.96 &  1343.97 &   1.68 &\nod &  45 \\
   490.16 &  1349.03 &   6.01 & 5.12 & L &    490.80 &  1349.39 &   6.18 &\nod &\nod \\
   490.46 &  1349.80 &   0.46 & 7.23 & R &    490.42 &  1350.17 &   0.46 &  57 &  44 \\
   490.93 &  1349.27 &   0.71 & 5.30 & R &    490.81 &  1349.62 &   0.81 &  58 &  45 \\
   532.53 &  1450.88 &   0.10 & 7.05 & R & \ndet                         &  59 &  46 \\
   548.39 &  1494.14 &   0.35 & 6.18 & R & \ndet                         &  60 &  48 \\
   551.21 &  1498.99 &   6.43 & 3.71 & L &    550.84 &  1499.18 &   5.39 &  61 &  49 \\
   555.60 &  1512.50 &   0.10 & 6.18 & R & \ndet                         &\nod &  50 \\
   602.73 & $-$97.60 &   0.10 & 3.19 & R & \ndet                         &  62 &  52 \\
   604.17 & $-$98.34 &   0.07 & 3.19 & L & \ndet                         &  62 &  52 \\
     \ndetb                          & L &    703.30 &$-$198.98 &   2.24 &\nod &\nod \\
   722.24 &$-$175.64 &   0.25 &16.20 & L & \ndet                         &\nod &\nod \\
   726.76 &$-$132.36 &   0.29&$-$0.86& L & \ndet                         &\nod &\nod \\
   730.11 &$-$199.21 &   0.98 &11.27 & R &    730.19 &$-$198.96 &   3.12 &\nod &\nod \\
   730.45 &$-$198.92 &   0.68 &11.27 & L & \ndet                         &\nod &\nod \\
     \ndetb                          & R &    737.67 &$-$130.51 &   1.56 &\nod &\nod \\
   737.75 &$-$128.49 &   2.91&$-$0.50& L &    739.15 &$-$127.99 &   1.35 &\nod &\nod \\
   738.05 &$-$128.53 &   1.60&$-$0.50& R &    739.13 &$-$128.07 &   0.73 &  72 &\nod \\
     \ndetb                          & R &    738.06 &$-$139.43 &   2.39 &\nod &\nod \\
     \ndetb                          & L &    738.45 &$-$141.27 &   3.31 &\nod &\nod \\
   740.38 &$-$152.14 &   0.16 &16.37 & L & \ndet                         &\nod &\nod \\
   741.08 &$-$147.91 &  43.21 & 1.96 & L &    741.17 &$-$148.09 &  44.99 &  70 &\nod \\
   741.42 &$-$147.68 &  24.28 & 1.78 & R &    741.41 &$-$147.77 &  23.53 &  70 &\nod \\
   741.95 &$-$152.95 &  67.91&$-$0.15& R &    741.71 &$-$152.78 &  58.47 &  71 &\nod \\
   742.00 &$-$152.89 & 279.86&$-$0.15& L &    741.72 &$-$152.74 & 213.96 &  71 &\nod \\
     \ndetb                          & L &    744.98 &$-$157.38 &   6.77 &\nod &\nod \\
     \ndetb                          & R &    745.24 &$-$157.51 &   1.66 &\nod &\nod \\
   745.28 &$-$135.81 &   2.10 & 0.02 & R &    745.50 &$-$135.51 &   2.14 &  73 &\nod \\
   754.08 & $-$81.73 &   1.94 & 7.93 & L &    754.02 & $-$81.97 &   5.75 &\nod &\nod \\
   754.36 & $-$82.29 &   0.13 & 7.93 & R &    753.88 & $-$81.95 &   0.60 &\nod &\nod \\
   791.08 &$-$240.26 &   0.89 & 5.82 & L &    791.92 &$-$240.58 &   0.60 &\nod &\nod \\
   792.01 &$-$240.73 &  20.79 & 5.82 & R &    791.90 &$-$241.28 &  16.50 &  82 &\nod \\
   801.73 &$-$134.88 &   0.32 & 3.36 & L & \ndet                         &\nod &\nod \\
   802.12 &$-$134.72 &   0.29 & 3.36 & R & \ndet                         &\nod &\nod \\
   813.86 &$-$137.85 &   0.11 & 3.54 & L & \ndet                         &\nod &\nod \\
   814.00 &$-$117.46 &   1.15 & 1.78 & R & \ndet                         &\nod &\nod \\
   817.15 &$-$121.02 &   0.07 & 3.54 & L & \ndet                         &\nod &\nod \\
   818.16 &$-$121.13 &   0.09 & 3.71 & R & \ndet                         &\nod &\nod \\
   844.63 &$-$1290.71&   0.10 &10.39 & R & \ndet                         &\nod &\nod \\
  1343.27 &    50.88 &   0.27 &11.45 & L & \ndet                         &  85 &\nod \\
  1343.29 &    50.62 &   0.24 &11.45 & R &   1342.93 &    51.59 &   0.28 &  85 &\nod \\
  1705.89 &$-$590.11 &   0.56 &13.73 & R &   1706.08 &$-$590.47 &   0.53 &  86 &  76 \\
  1784.55 &$-$585.60 &   0.14 &16.20 & L & \ndet                         &  88 &\nod \\
  1784.67 &$-$596.26 &   7.99 &11.98 & R &   1785.50 &$-$596.35 &   4.81 &  89 &  78 \\
  1784.89 &$-$596.30 &   0.54 &12.15 & L &   1785.14 &$-$596.65 &   0.33 &  89 &  78 \\
  1798.07 &$-$598.13 &   2.20 &11.98 & R &   1796.05 &$-$598.82 &   0.73 &\nod &  79 \\
  1812.47 &$-$609.78 &   0.22 &11.98 & R & \ndet                         &\nod &\nod \\
\hline
\end{tabular}
\end{minipage}
\end{table*}

\begin{table*}
\begin{minipage}{175mm}
\caption{Detected Masers at 1667~MHz}
\label{tab-1667}
\begin{tabular}{rrrrrrr}
\hline
\multicolumn{3}{c}{April 2008} & & & {2004} & {2000} \\
{RA Off.} &
{Dec Off.} &
{$S_\nu$} &
{$v_\mathrm{LSR}$} &
{} &
{Feature} &
{Feature} \\
{(mas)} &
{(mas)} &
{(Jy)} &
{(km\,s$^{-1}$)} &
{Pol} &
{Number} &
{Number} \\
\hline
      0.46 &   $-$1.29 &  0.10 &   10.00 & L &  91 &  81 \\
      0.68 &   $-$2.57 &  0.07 &   11.93 & R &  90 &  80 \\
    102.06 &    207.73 &  0.28 &   10.00 & L &  92 &  84\footnote{Blended with 83.} \\
    127.87 &  $-$56.33 &  0.32 &   12.63 & L &  93 &  85 \\
    127.91 &  $-$56.82 &  0.09 &   14.21 & R &\nod &\nod \\
    128.40 &  $-$55.55 &  0.14 &   12.63 & R &  93 &  85 \\
    202.46 &    708.84 &  1.11 &   10.00 & R &  98 &  92 \\
    207.05 &    674.33 &  2.29 &    5.43 & L & 101 &  93 \\
    207.27 &    674.04 &  0.35 &    8.07 & R & 102 &  94 \\
    220.16 &    617.02 &  0.38 &    5.26 & L & 104 &\nod \\
    220.59 &    616.61 &  0.09 &    8.07 & R & 103 &\nod \\
    221.97 &    655.89 &  3.30 &    5.96 & L & 105 &  97 \\
    232.91 &    560.79 &  0.13 &    6.49 & R &\nod &\nod \\
    233.17 &    560.93 &  0.43 &    6.31 & L &\nod &\nod \\
    233.56 &    551.78 &  0.42 &    6.49 & L & 107 &\nod \\
    233.84 &    553.89 &  0.11 &    9.12 & R & 106 &\nod \\
    234.24 &    594.13 &  0.39 &    9.47 & R &\nod &\nod \\
    235.04 &    586.21 &  0.07 &    6.66 & R &\nod &\nod \\
    458.15 &   1294.89 &  0.16 &    6.14 & R & 108 &  98 \\
    463.91 &   1299.16 &  0.07 &    5.79 & L & 109 &\nod \\
    491.80 &   1352.32 &  0.49 &    5.61 & L & 111 &  99 \\
    741.41 & $-$149.01 &  0.59 &    6.31 & L &\nod &\nod \\
    746.64 & $-$142.63 & 10.99 &    1.40 & L & 115 &\nod \\
    746.76 & $-$142.93 & 25.11 &    1.22 & R & 115 &\nod \\
    746.82 & $-$160.84 & 44.10 &    5.08 & L &\nod &\nod \\
    746.90 & $-$161.32 & 13.06 &    4.91 & R &\nod &\nod \\
    755.39 & $-$159.48 &  0.16 &    2.63 & L &\nod &\nod \\
    756.52 & $-$148.27 &  0.16 & $-$0.01 & L &\nod &\nod \\
    777.82 & $-$110.83 &  0.75 &    6.49 & L &\nod &\nod \\
    779.13 & $-$111.74 &  8.34 &    0.52 & R &\nod &\nod \\
    795.04 &  $-$28.92 &  0.07 &   10.70 & L &\nod &\nod \\
    804.93 & $-$111.44 &  0.19 &    2.98 & R &\nod &\nod \\
    811.13 & $-$115.66 &  0.22 &    3.33 & R &\nod &\nod \\
    812.19 & $-$112.55 &  0.17 &    4.38 & R &\nod &\nod \\
   1781.81 & $-$583.39 &  0.34 &   15.44 & L & 120 & 116 \\
   1783.87 & $-$584.66 &  0.27 &   12.98 & R & 121 & 117 \\
   1826.12 & $-$588.85 &  0.16 &   12.63 & R &\nod & 118 \\
   1833.13 & $-$561.60 &  0.26 &   14.74 & L & 122 & 119 \\
   1833.43 & $-$561.64 &  1.07 &   12.46 & R & 123 & 120 \\
\hline
\end{tabular}
\end{minipage}
\end{table*}

\begin{table*}
\begin{minipage}{175mm}
\caption{Detected Masers at 1720~MHz with the VLBA}
\label{tab-1720}
\begin{tabular}{rrrrrrrr}
\hline
\multicolumn{3}{c}{April 2008} &
&
&
\multicolumn{3}{c}{February 2008}
\\
{RA Off.\footnote{The 1720-MHz masers have been put into the 1665/1667-MHz
reference frame using the absolute position of the 1720-MHz reference feature
and the coordinates of spot A in \citet{slysh2010} as described in
Section~\ref{alignment}.}} &
{Dec Off.} &
{$S_\nu$} &
{$v_\mathrm{LSR}$} &
{} &
{RA Off.} &
{Dec Off.} &
{$S_\nu$} 
\\
{(mas)} &
{(mas)} &
{(Jy)} &
{(km\,s$^{-1}$)} &
{Pol} &
{(mas)} &
{(mas)} &
{(Jy)} \\
\hline
  110.05  &    234.34  &    2.87 & 9.83 & L &   110.03  &   233.92 &   1.52 \\
  110.22  &    234.46  &    0.64 &11.02 & R &   110.24  &   234.49 &   0.61 \\
  125.45  &    202.79  &    0.79 &11.19 & R & \ndet                         \\
  125.60  &    202.94  &    0.68 &10.17 & L &   125.66  &   203.27 &   0.39 \\
\ndet                            & 7.96 & R &   177.28  &   462.60 &   0.32 \\
  177.65  &    463.27  &    5.59 & 6.94 & L &   177.98  &   463.92 &   8.25 \\
  178.82  &    471.93  &    3.28 & 9.15 & R & \ndet                         \\ 
\ndet                            & 8.98 & L &   182.58  &   480.81 &   2.07 \\
  184.11  &    353.33  &    0.31 & 8.81 & L & \ndet                         \\
  184.12  &    353.95  &    1.66 & 9.66 & R &   184.57  &   354.94 &   1.64 \\
  185.22  &    495.08  &   88.55 & 8.47 & R & \ndet\footnote{Blended with the next spot.} \\
  185.42  &    488.00  &  301.46 & 8.47\footnote{Based on the MERLIN observations, the position of this maser is 20:38:36.425 +42:37:34.73 (J2000) with a positional uncertainty of 20~mas.} & R &   185.419 &   488.000& 439.71 \\
  185.68  &    488.04  &   81.73 & 7.28 & L &   185.28  &   487.62 &  66.81 \\
  185.60  &    499.28  &    9.74 & 7.79 & L &   185.84  &   499.40 &   7.76 \\
\ndet                            & 9.66 & L &   186.29  &   489.70 &  14.14 \\
  189.55  &    376.55  &    2.75 &11.02 & L &   189.89  &   376.70 &   1.80 \\
  192.82  &    387.85  &    0.44 &10.51 & L &   193.96  &   390.29 &   0.51 \\
  197.60  &    441.05  &    0.42 & 9.15 & L & \ndet                         \\
  210.36  &    391.62  &   24.26 & 8.98 & L &   210.38  &   391.72 &   8.64 \\
\ndet                            & 8.13 & R &   210.38  &   143.35 &   0.91 \\
  210.39  &    391.99  &    2.22 &10.00 & R &   210.43  &   391.62 &   4.28 \\
  211.39  &    405.84  &    2.18 & 9.83 & R & \ndet                         \\
  214.02  &    400.60  &   64.37 & 9.66 & L &   214.20  &   400.58 &  66.46 \\
  217.03  &    433.22  &    0.72 & 9.83 & R & \ndet                         \\
  217.24  &    433.86  &    1.16 & 8.81 & L &   216.53  &   433.30 &   0.74 \\
  219.51  &    152.90  &    0.35 & 7.79 & R &   219.90  &   152.46 &   0.34 \\
  221.52  &    403.23  &    6.34 & 9.66 & L & \ndet                         \\
  223.61  &    404.14  &    2.78 & 9.66 & R &   224.14  &   404.83 &   2.16 \\
\hline
\end{tabular}
\end{minipage}
\end{table*}

\label{lastpage}

\end{document}